\documentstyle{article}

\input{psfig}

\begin{document}

\title{Measurements of Charged Current Reactions of $\nu_e$ on $^{12}C$}

\author{
L.B. Auerbach,$^8$ R.L. Burman,$^5$ D.O. Caldwell,$^3$ E.D. Church,$^1$ \\
J.B. Donahue,$^5$ A. Fazely,$^7$ G.T. Garvey,$^5$ R.M. Gunasingha,$^7$ 
R. Imlay,$^6$ \\
W.C. Louis,$^5$ R. Majkic,$^{8}$ A. Malik,$^6$ W. Metcalf,$^6$ 
G.B. Mills,$^5$ \\
V. Sandberg,$^5$ D. Smith,$^4$ 
I. Stancu,$^1$\footnote{now at University of Alabama, Tuscaloosa, AL 35487}
M. Sung,$^6$ 
R. Tayloe,$^5$\footnote{now at Indiana University, Bloomington, IN 47405} \\ 
G.J. VanDalen,$^1$ 
W. Vernon,$^2$ N. Wadia,$^6$ D.H. White,$^5$ S. Yellin$^3$\\
(LSND Collaboration) \\
$^1$ University of California, Riverside, CA 92521 \\
$^2$ University of California, San Diego, CA 92093 \\
$^3$ University of California, Santa Barbara, CA 93106 \\
$^4$ Embry Riddle Aeronautical University, Prescott, AZ 86301 \\
$^5$ Los Alamos National Laboratory, Los Alamos, NM 87545 \\
$^6$ Louisiana State University, Baton Rouge, LA 70803 \\
$^7$ Southern University, Baton Rouge, LA 70813 \\
$^8$ Temple University, Philadelphia, PA 19122 }

\date{\today}
\maketitle

\begin{abstract}
Charged Current reactions of $\nu_e$ on $^{12}C$ have been studied 
using a $\mu^+$ decay-at-rest $\nu_e$ beam 
at the Los Alamos Neutron Science Center.
The cross section for the exclusive 
reaction $^{12}C(\nu_e,e^-)^{12}N_{g.s.}$ 
was measured to be $(8.9\pm0.3\pm0.9)\times10^{-42}$ cm$^2$.
The observed energy dependence of the cross section and 
angular distribution of the outgoing electron agree well 
with theoretical expectations.
Measurements are also presented for inclusive transitions 
to $^{12}N$ excited states, $^{12}C(\nu_e,e^-)^{12}N^*$ and compared 
with theoretical expectations.
The measured cross section, $(4.3\pm0.4\pm0.6)\times10^{-42}$ cm$^2$, 
is somewhat lower than previous measurements and 
than a continuum random phase approximation calculation.
It is in better agreement with a recent shell model calculation.
\end{abstract}

\section{Introduction}
\label{sec:intro}

In recent years neutrino interactions with nuclear targets have been used 
to detect low energy neutrinos ($<500$ MeV) from many sources: 
solar, atmospheric, supernova explosions, reactors and accelerators.
An understanding of the nuclear cross sections is necessary 
for interpretation of the measurements.
Especially important nuclei are $^{12}C$ and $^{16}O$ 
because of the widespread use of oil and water detectors.

Many calculation techniques have been used to determine neutrino-nuclear 
cross sections.
Shell model techniques work best at lower energies 
where transitions to continuum states are not large.
At intermediate energies the Continuum Random Phase Approximation (CRPA) 
is frequently used, while at still higher energies 
the Fermi gas model is thought to work well.
Comparison of different calculations of the cross section for 
a particular process can provide an indication 
of the uncertainty involved.
Experimental measurements of some cross sections are, however, necessary 
to establish the range of validity of 
the different calculation techniques.

Relatively few measurements of neutrino-nucleus cross sections exist in 
the energy region of the present experiment, $E_\nu<52$.8 MeV. 
The best measured nucleus is carbon for which three experiments, 
including the Liquid Scintillator Neutrino Detector (LSND) experiment, 
have previously reported results.  
E225\cite{E225} at LAMPF, the KARMEN Collaboration\cite{KARMEN} 
at the ISIS facility of the Rutherford Laboratory and 
LSND\cite{LSND97a} have measured the cross section for the exclusive 
reaction $^{12}C(\nu_e,e^-)^{12}N_{g.s.}$ and for the inclusive 
reaction $^{12}C(\nu_e,e^-)^{12}N^*$ to all other accessible $^{12}N$ 
final states.
The $^{12}N$ ground state reaction dominates the total yield 
as it is the only allowed $(l=0)$ transition that occurs in this process. 
The cross section for producing the $^{12}N$ ground state can be 
calculated to an accuracy of approximately $5\%$ 
as it can be represented in terms of form factors\cite{Fukugita88} 
that can be reliably extracted from other measurements.
Calculation of the inclusive cross section for transitions to 
excited states of $^{12}N$ is much less straightforward.
Various theoretical techniques, each with their own strengths 
and limitations, have been used to calculate the cross 
section\cite{Donnelly,Kolbe94,Kolbe99,Volpe00,Hayes00,Auerbach97,Singh98}.
Comparison with measurements may help clarify the theoretical picture.
In this paper we report our final results for these processes, 
including measurements of the angular distribution
of the electron with respect to the $\nu_e$ direction 
and the energy dependence of the ground state transition.

Measurements also exist for two processes closely related 
to $\nu_e$ carbon scattering: $\mu^-$ capture on $^{12}C$\cite{Suzuki87} 
and $\nu_\mu$ scattering on carbon using a beam of $\nu_\mu$ 
from $\pi^+$ decay-in-flight (DIF)\cite{LSND97b}.
Because these three processes occur at different energies, $E_\nu$, and 
momentum transfers, $Q$, 
they constrain different aspects of theoretical calculations.
A good test of a theoretical procedure is its ability 
to predict all three processes.
For the $\nu_e$ carbon measurement $E_\nu\approx32$ MeV, 
$Q\approx50$ MeV/$c$ and the inclusive cross section is dominated 
by transitions to low multipoles ($1^+,1^-,2^-$).
In contrast, for the $\nu_\mu$ carbon measurement $E_\nu\approx180$ MeV, 
$Q\approx200$ MeV/$c$ and excitations occur up to 100 MeV.
The $\mu^-$ capture process, which occurs from the $S$ state, 
is intermediate between these two processes with $Q\approx90$ MeV/$c$.

The measurement\cite{LSND97b} of the inclusive cross section 
for $^{12}C(\nu_\mu,\mu^-)^{12}N^*$ several years ago by LSND attracted 
substantial interest because a CRPA calculation\cite{Kolbe94} predicted 
a cross section almost twice as large as that observed.
An improved calculation by the same group\cite{Kolbe99} 
together with an improved calculation of the neutrino energy spectrum 
and flux has reduced but not eliminated the discrepancy 
with the measured cross section.
Recent calculations using the shell model\cite{Volpe00,Hayes00} are 
in better agreement with the measured cross section.

Hayes and Towner\cite{Hayes00} calculated a cross section 
of $4.1\times10^{-42}$ cm$^2$ for the process $^{12}C(\nu_e,e^-)^{12}N^*$ 
using the same shell model procedure that provided the best agreement 
with data for the reaction $^{12}C(\nu_\mu,\mu^-)^{12}N^*$.
This is lower than both the earlier\cite{Kolbe94} and 
the more recent\cite{Kolbe99} CRPA calculations by Kolbe {\it et al.} 
of $6.3\times10^{-42}$ cm$^2$ and $5.5\times10^{-42}$ cm$^2$ respectively.
Thus measurements of this cross section can provide a useful test 
of the relative merits of the different theoretical techniques 
that have been used.

\section{The Neutrino Source}
\label{sec:source}

\begin{figure}
\centerline{\psfig{figure=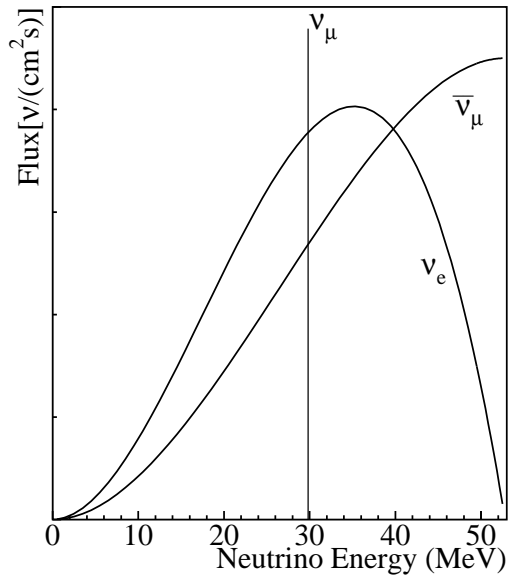,height=3.3in}}
\caption{Flux shape of neutrinos from pion and muon decay at rest.}
\label{fig:flux}
\end{figure}

The data reported here were obtained between 1994 and 1997 
at the Los Alamos Neutron Science Center (LANSCE)
using neutrinos produced at the A6 proton beam stop.
We chose to exclude data obtained in 1998 from this analysis 
because only electrons with reconstructed energies above 20.4 MeV 
were fully processed for that year's data.
Since this analysis is dominated by systematics 
we decided to limit our analysis to the 1994-1997 data 
which has a uniform efficiency above 18 MeV.
The neutrino source is described in detail elsewhere\cite{LSND_NIM}.
In 1994 and 1995 the beam stop consisted of a 30 cm water target 
surrounded by steel shielding and followed by a copper beam dump.
The high-intensity 798 MeV proton beam from the linear accelerator 
generated a large pion flux from the water target.
The flux of $\nu_e$ used for the measurements reported here arise 
from the decay at rest (DAR) of stopped $\pi^+$ and $\mu^+$.
This decay chain yields almost equal intensities 
of $\nu_e,~\bar{\nu}_\mu$ and $\nu_\mu$ with the well-determined energy 
spectra shown in Fig. \ref{fig:flux}.

After the 1995 run the beam stop was substantially modified 
for accelerator production of tritium (APT) tests.
The most significant change for the analysis presented in this paper 
was the replacement of the water target by tungsten and 
other materials with high atomic number.
This resulted in reduced $\pi^+$ production and a lower DAR neutrino 
flux, largely due to the change in the neutron to proton ratio 
in the target.

The corresponding decay chain for $\pi^-$ and $\mu^-$ is 
highly suppressed due to three factors. 
First, production of $\pi^-$ is smaller than for $\pi^+$. 
Second, $\pi^-$ which stop are absorbed by nuclear interactions. 
Finally, most $\mu^-$ which stop are absorbed before they can decay. 
These stopped $\mu^-$ arise from $\pi^-$ which decay in flight.

A few percent of the $\pi^+~(\pi^-)$ produced at the beam dump 
decay in flight to $\nu_\mu~(\bar{\nu}_\mu)$ with energies up to 300 MeV.
Those $\nu_\mu$ above muon production threshold provide the source used 
for our measurement of $^{12}C(\nu_\mu,\mu^-)^{12}N^*$\cite{LSND97b}.
In the analysis of the reaction $^{12}C(\nu_e,e^-)^{12}N^*$ presented in 
section \ref{sec:exstate} below we correct for a small background arising 
from $\mu^\pm$ produced by high energy $\nu_\mu$ and $\bar{\nu}_\mu$.

The LANSCE beam dump has been used as the neutrino source for previous 
experiments\cite{Willis80,Krak92,Free93}.
A calibration experiment\cite{All89} measured the rate of stopped $\mu^+$ 
from a low-intensity proton beam incident on an instrumented beam stop.
The rate of stopped $\mu^+$ per incident proton was measured 
as a function of several variables and 
used to fine-tune a beam dump simulation program\cite{Bur90}.
The simulation program can then be used to calculate the flux 
for any particular beam dump configuration.
The calibration experiment determined the DAR flux to $\pm7\%$ 
for the proton energies and beam stop configurations used at LANSCE.
This uncertainty provides the largest source of systematic error for 
the cross sections presented here.
The LANSCE proton beam typically had a current of 800 $\mu$A 
at the beam stop during the 1994-1995 running period and 1000 $\mu$A 
for 1996-1997. 
For 1994 and 1995 the energy was approximately 770 MeV at the beam stop 
due to energy loss in upstream targets, 
while it was approximately 800 MeV in 1996 and 1997.
The water target was out for $32\%$ of the 1995 data.
Upstream targets contributed 1.4$\%$ to the DAR flux in 1994 and 1995.
The DAR $\nu_e$ flux averaged over the LSND detector was 
$3.08\times10^{13}$ cm$^{-2}$ for 1994 and $3.45\times10^{13}$ cm$^{-2}$ 
for 1995.

The 1996-1998 data were obtained with the APT beam stop.
There were no upstream targets for almost all of the data taking 
for this period.
The DAR $\nu_e$ flux averaged over the LSND detector was 
$1.32\times10^{13}$ cm$^{-2}$ for 1996 and  $2.73\times10^{13}$ cm$^{-2}$ 
for 1997. 
For the full data sample used in this paper the $\nu_e$ flux 
is $10.58\times10^{13}$ cm$^{-2}$.

\section{The LSND Detector}
\label{sec:lsnd}

\begin{figure}
\centerline{\psfig{figure=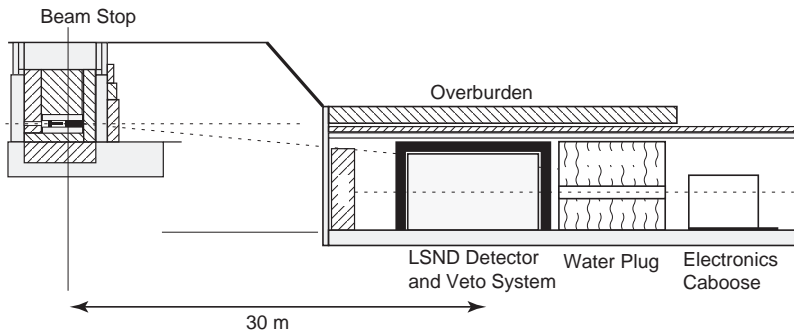,height=1.7in}}
\caption{Detector enclosure and target area configuration, elevation view.}
\label{fig:detector}
\end{figure}

The detector is located 29.8 m downstream of the proton beam stop 
at an angle of $12^\circ$ to the proton beam.  
Figure \ref{fig:detector} shows a side-view of the setup.  
Approximately 2000 g/cm$^2$ of shielding above the detector attenuates 
the hadronic component of cosmic rays to a negligible level.  
The detector is also well shielded from the beam stop 
so that beam associated neutrons are attenuated to a negligible level.  
Enclosing the detector, except on the bottom, 
is a highly efficient liquid scintillator veto shield 
which is essential to reduce contributions 
from the cosmic ray muon background to a low level.    
Reference \cite{LSND_NIM} provides a detailed description of the detector,
veto, and data acquisition system which we briefly review here.  

The detector is an approximately cylindrical tank containing 167 metric 
tons of liquid scintillator and viewed by 1220 uniformly spaced $8''$ 
Hamamatsu photomultiplier tubes (PMT) covering $25\%$ of the surface 
inside the tank wall. 
When the deposited energy in the tank exceeds a threshold of 
approximately 4 MeV electron-equivalent energy and 
there are fewer than 4 PMT hits in the veto shield, 
the digitized time and pulse height of each of these PMTs 
(and of each of the 292 veto shield PMTs) are recorded.  
A veto, imposed for 15.2 $\mu$s following the firing of $>5$ veto PMTs, 
substantially reduces ($10^{-3}$) 
the large number of background events arising from the decay 
of cosmic ray muons that stop in the detector.  
Activity in the detector or veto shield during the 51.2 $\mu$s 
preceding a primary trigger is also recorded, 
provided there are $>17$ detector PMT hits or $>5$ veto PMT hits.  
This activity information is used in the analysis to reject events 
arising from muon decay. 
Data after the primary event are recorded for 1 ms with a threshold 
of 21 PMTs (approximately 0.7 MeV electron-equivalent energy). 
This low threshold is used for the detection of 2.2 MeV $\gamma$'s 
from neutron capture on free protons.  
The processes measured in this paper, $^{12}C(\nu_e,e^-)^{12}N_{g.s.}$ 
and $^{12}C(\nu_e,e^-)^{12}N^*$, do not produce neutrons. 
Thus, in the present analysis, detection of 2.2 MeV $\gamma$'s 
is used to help determine beam-related backgrounds 
with associated neutrons.
The detector operates without reference to the beam spill, 
but the state of the beam is recorded with the event.  
Approximately $94\%$ of the data is taken between beam spills.  
This allows an accurate measurement and subtraction 
of cosmic ray background surviving the event selection criteria.  

Most triggers due to electrons from 
stopped muon decays (Michel electrons) are identified 
by a preceding activity produced by the decay muon.  
Occasionally, however the muon will not satisfy 
the activity threshold of $>17$ detector PMT hits or $>5$ veto PMT hits.  
For several LSND analyses, including the present one, 
it is desirable to further reduce the number 
of unidentified Michel electrons. 
Therefore, for data recorded after 1994 all PMT information was recorded 
for a period of 6 $\mu$s (2.7 muon lifetimes) 
preceding certain primary events.
For the present analysis this ``lookback" information is used 
to further reduce the cosmic ray muon background 
as described in Section \ref{sec:electron}. 

The detector scintillator consists of mineral oil ($CH_2$) in which is 
dissolved a small concentration (0.031 g/l) of b-PBD\cite{Ree93}. 
This mixture allows the separation of \v{C}erenkov light and 
scintillation light and produces about 33 photoelectrons per MeV of 
electron energy deposited in the oil.  
The combination of the two sources of light provides direction 
information for relativistic particles and makes 
particle identification (PID) possible.
Note that the oil consists almost entirely of carbon and hydrogen.  
Isotopically the carbon is $1.1\%~^{13}C$ and $98.9\%~^{12}C$.

The veto shield encloses the detector on all sides except the bottom.  
Additional counters were placed below the veto shield 
after the 1993 run to reduce cosmic ray background entering 
through the bottom support structure.  
More counters were added after the 1995 run.
The main veto shield\cite{Nap89} consists of a 15-cm layer of liquid 
scintillator in an external tank and 15 cm of lead shot 
in an internal tank.  
This combination of active and passive shielding tags cosmic ray muons 
that stop in the lead shot.  
A veto inefficiency $<10^{-5}$ is achieved with this detector 
for incident charged particles.  

\section{Analysis Techniques}
\label{sec:analysis}

Each event is reconstructed using the hit time and pulse height 
of all hit PMTs in the detector\cite{LSND_NIM}.
The present analysis relies on the reconstructed energy, position, 
direction, and two PID parameters, $\chi^\prime_{tot}$ and $\alpha$, 
as described later in this section.
The particle direction is determined from the \v{C}erenkov cone.
The parameters $\chi^\prime_{tot}$ and $\alpha$ are used to distinguish 
electron events from events arising from interactions of 
cosmic ray neutrons in the detector. 
We directly measure the response of the detector to electrons and 
neutrons in the energy range of interest for this analysis 
by using copious control data samples.
We also use a GEANT Monte Carlo simulation, LSNDMC\cite{McI95} to 
simulate events  in the detector.

The response of the detector to electrons was determined from a large, 
essentially pure sample of electrons (and positrons) from the decay of 
stopped cosmic ray $\mu^\pm$ in the detector.
The known energy spectra for electrons from muon decay was used 
to determine the absolute energy calibration, 
including its small variation over the volume of the detector.
The energy resolution was determined from the shape of the electron 
energy spectrum and was found to be $6.6\%$ at the 52.8 MeV end-point.

There are no tracking devices in the LSND detector. 
Thus, event positions must be determined solely from the PMT information.
The reconstruction process determines an event position by minimizing 
a function $\chi_r$ which is based on the time of each PMT hit corrected 
for the travel time of light from the assumed event position 
to the PMT\cite{LSND_NIM}.
The procedure used in several previous analyses systematically shifted 
event positions away from the center of the detector and thus 
effectively reduced the fiducial volume\cite{At96}.
The reconstruction procedure has been analyzed in detail and 
an improved reconstruction procedure was developed which reduces this 
systematic shift and provides substantially better position resolution.
This procedure also provides results which agree well with positions 
obtained from the event likelihood procedure described 
in Ref. \cite{At98}.
In the analysis presented in this paper, a fiducial cut is imposed 
by requiring $D>35$ cm, where $D$ is the distance between 
the reconstructed event position and the surface tangent 
to the faces of the PMTs. 
Events near the bottom of the detector ($y<-120$ cm) are also removed, 
as discussed in Section \ref{sec:electron}.

The particle identification procedure is designed to separate particles 
with velocities well above \v{C}erenkov threshold from particles 
below \v{C}erenkov threshold.
The procedure makes use of the four parameters defined 
in Ref. \cite{LSND_NIM}.
Briefly, $\chi_r$ and $\chi_a$ are the quantities minimized for 
the determination of the event position and direction, $\chi_t$ is 
the fraction of PMT hits that occur more than 12 ns after the fitted 
event time and $\chi_{tot}$ is proportional to the product of $\chi_r$, 
$\chi_a$ and $\chi_t$.

Several previous LSND analyses \cite{LSND97a,LSND97b,At96} have 
used $\chi_{tot}$ for particle identification.
The distribution of $\chi_{tot}$ for electrons, however, has 
a small variation with electron energy and with the position of the event.
Therefore, in this paper, we used a modified 
variable, $\chi^\prime_{tot}$, with a mean of zero and sigma of one, 
independent of the electron energy and positions. 
We also used the variable, $\alpha$, which is based on the event 
likelihood procedures discussed in Ref. \cite{At98}.
As in \cite{At98}, $\alpha$ comes from a separate event reconstruction 
than that which produced $\chi^\prime_{tot}$.
It is similar to the parameter $\rho$ discussed there, 
which is based on the ratio of \v{C}erenkov to scintillator light.
The $\alpha$ parameter varies from 0 to 1 and 
peaks at one for electrons and at 0.3 for neutrons.
The combination $\chi_\alpha=\chi^\prime_{tot}+10(1-\alpha)$ provides 
better separation of electrons and neutrons than $\chi^\prime_{tot}$ 
by itself.

\begin{figure}
\centerline{\psfig{figure=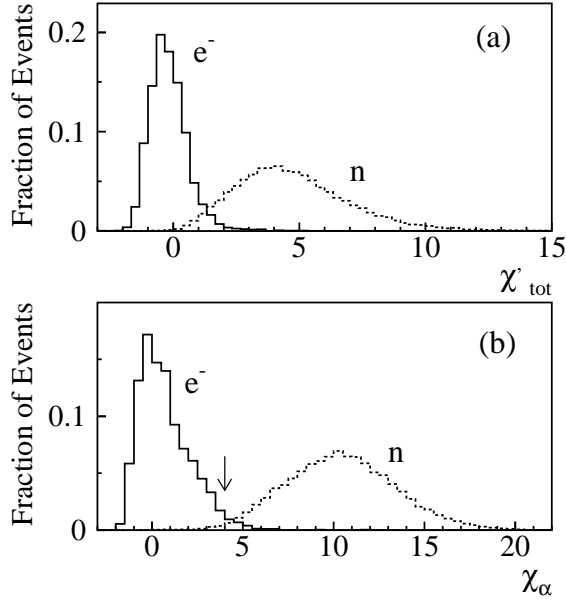,height=3.3in}}
\caption{Particle identification parameters (a) $\chi^\prime_{tot}$ and 
(b) $\chi_\alpha$ for electrons and neutrons.
In the present analysis we require $\chi_\alpha<4.0$ as indicated by the 
arrow in (b).}
\label{fig:eid}
\end{figure}
Figure \ref{fig:eid}(a) shows the $\chi^\prime_{tot}$ distribution 
for electrons from stopping $\mu$ decay and for cosmic ray neutrons 
with electron equivalent energies in the $18<E_e<50$ MeV range. 
Neutrons, after thermalizing, undergo a capture reaction, 
$n+p\rightarrow d+\gamma$. 
The 2.2 MeV $\gamma$'s are used to select a clean sample 
of neutron events. 
For a neutron $E_e$ is the equivalent electron energy corresponding to 
the observed total charge.
Figure \ref{fig:eid}(b) shows the $\chi_\alpha$ distribution for 
the same events.
In the present analysis we eliminate most cosmic ray neutron background 
by requiring $\chi_\alpha<4.0$. 
We note that a modest particle identification requirement was imposed 
in the initial data processing that created the samples analyzed here. 
The effect of this requirement is also included in the analysis.

Beam-off data taken between beam spills play a crucial role 
in the analysis of this experiment.
Most event selection criteria are designed to reduce 
the cosmic ray background while retaining high acceptance 
for the neutrino process of interest.
Cosmic ray background which remains after all selection criteria 
have been applied is well measured with the beam-off data and 
subtracted using the duty ratio, the ratio of beam-on time 
to beam-off time.
The subtraction was performed separately for each year's data 
using the measured duty ratio for that year.
The ratio averaged over the full data sample was 0.0632.
Beam-on and beam-off data have been compared to determine if there are 
any differences other than those arising from neutrino interactions.
Any differences are small and the $1.1\%$ uncertainty in the duty ratio 
introduces a negligible effect in the present analysis.

\section{Electron Selection Criteria}
\label{sec:electron}

In this section we describe the selection criteria used 
to obtain a clean sample of inclusive electrons arising 
from neutrino interactions in the detector.
In the next section we present the analysis of the relatively pure sample 
of events from the process $^{12}C(\nu_e,e^-)^{12}N_{g.s.}$ 
which we obtain by requiring the detection of the positron 
from the $\beta$-decay of the $^{12}N_{g.s.}$.
Section \ref{sec:exstate}  then presents the analysis 
of the reaction $^{12}C(\nu_e,e^-)^{12}N^*$ 
using the sample of inclusive events 
without an identified positron from the $\beta$-decay of $^{12}N_{g.s.}$

A lower limit of 18.0 MeV is imposed on the electron energy 
to eliminate the large cosmic ray background from $^{12}B~\beta$-decay 
as well as most 15.1 MeV gamma rays from the neutral current excitation 
of carbon.
The $^{12}B$ nuclei arise from the absorption of stopped $\mu^-$ 
on $^{12}C$ nuclei in the detector.
The scattered electron from the reaction $^{12}C(\nu_e,e^-)^{12}N_{g.s.}$ 
has a maximum kinetic energy of 35.5 MeV due to the $Q$ value of 17.3 MeV.
Allowing for energy resolution we impose an upper limit of 40 MeV 
on the electron energy.
\begin{table}
\centering
\caption{The electron selection criteria and corresponding efficiencies 
for events with 18 MeV$<E_e<40$ MeV.}
\begin{tabular}{ccc}
\hline
   Quantity            &    Criteria           &   Efficiency    \\    
\hline
Fiducial volume        & $D>35$ cm,            & 0.880$\pm$0.055 \\
                       & $y>-120$ cm           &                 \\
Particle ID            & $\chi_\alpha<4$       & 0.940$\pm$0.018 \\
In-time veto           & $<4$ PMTs             & 0.988$\pm$0.010 \\
Past activity          & See text              & 0.635$\pm$0.012 \\
Future activity        & $\Delta t_f>8.8~\mu$s & 0.991$\pm$0.003 \\
Lookback               & likelihood            & 0.994$\pm$0.004 \\
DAQ and tape dead time &   --                  & 0.962$\pm$0.010 \\
\hline
Total                  &                       & 0.492$\pm$0.035 \\
\hline
\end{tabular}
\label{ta:electron}
\end{table}

The selection criteria and corresponding efficiencies for electrons 
with 18 MeV$<E_e<$ 40 MeV are shown in Table \ref{ta:electron}. 
The reconstructed electron position is required to be a 
distance $D>35$ cm from the surface tangent to the faces of the PMTs.
The requirement $y>-120$ cm removes a small region at the bottom 
of the detector for which the cosmic ray background is relatively high 
due to the absence of a veto below the detector.
There are $3.34\times10^{30}~^{12}C$ nuclei within this fiducial volume.
The fiducial volume efficiency, defined to be the ratio of the number 
of events reconstructed within the fiducial volume to the actual number 
within this volume, was determined to be $0.880\pm0.055$.
This efficiency is less than one because there is a systematic shift 
of reconstructed event positions away from the center of the detector 
as discussed in Section \ref{sec:analysis}.

Several selection criteria are designed to further reject 
cosmic ray induced events.
Events with more than three veto PMT hits or any bottom counter coincidence 
during the 500 ns event window are eliminated.
The past activity cut is designed to reject most electron events arising 
from cosmic ray muons which stop in the detector and decay.
This background has a time dependence given by the 2.2 $\mu$s muon lifetime.
The past activity selection criteria reject all events with activity 
within the past 20 $\mu$s with $>5$ veto PMT hits or $>17$ detector PMT hits.
We also reject any event with a past activity within 51 $\mu s$ 
with $>5$ veto PMT hits and $>500$ detector PMT hits. 
A small ($0.5\%$) loss of efficiency arises from a cut (made during initial 
data processing) on past activities that are spatially correlated 
with the primary event, within 30 $\mu$s of the primary event and 
have $\ge4$ veto PMT hits.

Muons which are misidentified as electrons are removed by requiring 
that there be no future activity consistent with a Michel electron.
Any event with a future activity with fewer than 4 veto PMT hits and 
more than 50 detector PMT hits within 8.8 $\mu$s is rejected.

Cosmic ray muons which fire $<6$ veto PMTs ($10^{-3}$ probability) and 
stop in the iron walls of the detector will not register 
as past activities.
Some of the decay electrons will radiate photons which will enter 
the detector and be reconstructed as electrons within the fiducial volume.
In previous analyses we simply relied 
on the beam-off subtraction procedure to remove this background.
Here we use the ``lookback'' information described 
in Section \ref{sec:lsnd} to reject events from this source.
This results in slightly smaller statistical errors 
in the final beam excess sample.

\begin{figure}
\centerline{\psfig{figure=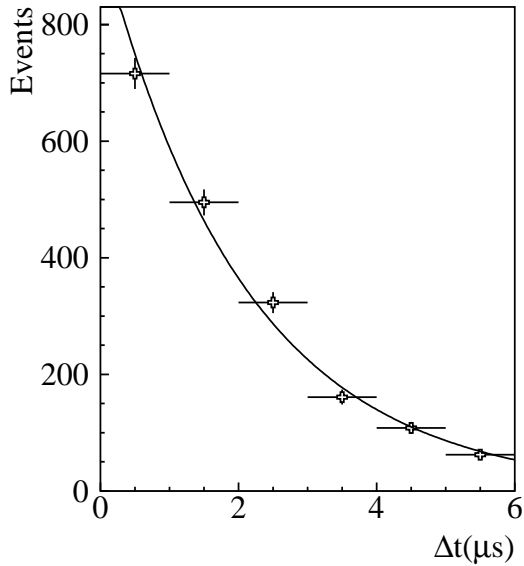,height=3.3in}}
\caption{The distribution of time between the primary and the veto signal for 
beam-off events rejected using the ``lookback'' information compared 
with a curve corresponding to the muon lifetime.}
\label{fig:beta5}
\end{figure}

For primary events with $>300$ PMT hits and no activity 
within the past $35\mu$s ($20\mu$s) for 1995 data (1996-1998 data),
we recorded all PMT information for the 6 $\mu$s interval 
proceeding the event.
Muons with $<6$ veto PMT hits will appear in this ``lookback'' interval 
as a cluster of veto PMT hits spatially correlated with the primary event.
The distribution of time between the veto signals and the primary event 
should be consistent with the muon lifetime, 
and the distributions of veto PMT hits and veto pulse height 
should be consistent with that measured for muons 
producing $<6$ veto PMT hits.
We developed a likelihood procedure based on these distributions 
which allowed us to reduce the beam off background by $9\%$ 
with only a $0.6\%$ loss of efficiency for neutrino events\cite{Wad98}.
Figure \ref{fig:beta5} shows the time between the veto signal and 
the primary for rejected events.
The fitted lifetime of 2.08$\pm$0.07 $\mu$s agrees well with 
the expected average lifetime of 2.12 $\mu$s for stopping $\mu^+$ 
and $\mu^-$ in oil.

The acceptances for the past activity, the future activity, 
the ``lookback'' and the in-time veto cuts are obtained by applying these cuts 
to a large sample of random events triggered with the laser used 
for detector calibration.
These laser events are spread uniformly through the run and 
thus average over the small variation in run conditions.
The acceptance for the 15.1 $\mu$s trigger veto is included 
in the past activity efficiency.

\begin{figure}
\centerline{\psfig{figure=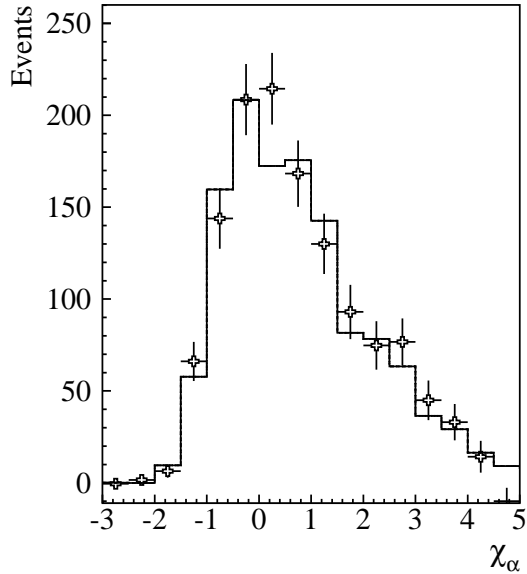,height=3.3in}}
\caption{The $\chi_\alpha$ distribution of the beam excess inclusive 
electron sample. 
The histogram shows the $\chi_\alpha$ distribution of Michel electrons.}
\label{fig:chialpha}
\end{figure}

A sample of Michel electrons 
was analyzed to obtain the acceptance of electrons for the PID cut.
Figure \ref{fig:chialpha} compares the $\chi_\alpha$ distribution of 
the inclusive electron sample with a Michel electron sample.
The agreement is excellent.
To eliminate any energy dependence, 
the Michel electrons are given weights 
as a function of energy so that the weighted spectrum agrees with 
the energy spectrum of electrons 
from the reaction $^{12}C(\nu_e,e^-)^{12}N_{g.s.}$.
The acceptance, however, is very insensitive 
to the assumed energy spectrum.
The beam excess distribution shown in Figure \ref{fig:chialpha} 
is obtained by subtracting the beam off distribution 
from the beam on distribution as discussed in Section \ref{sec:lsnd}.

\section{The Transition to the $^{12}N$ Ground State}
\label{sec:gstate}

\begin{figure}
\centerline{\psfig{figure=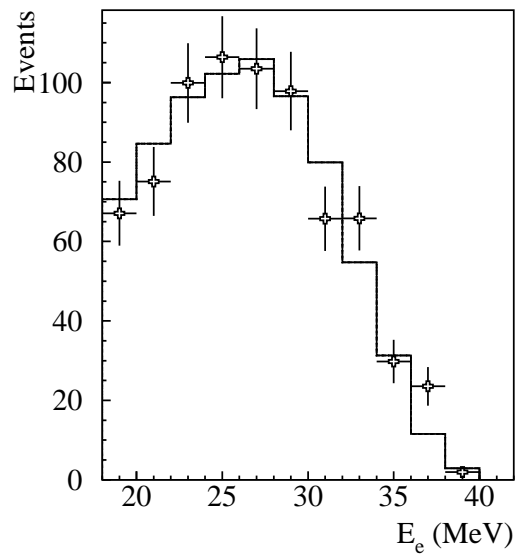,height=3.3in}}
\caption{The observed and expected (solid line) energy distributions 
for electrons from $^{12}C(\nu_e,e^-)^{12}N_{g.s.}$.}
\label{fig:eelec}
\end{figure}

\begin{figure}
\centerline{\psfig{figure=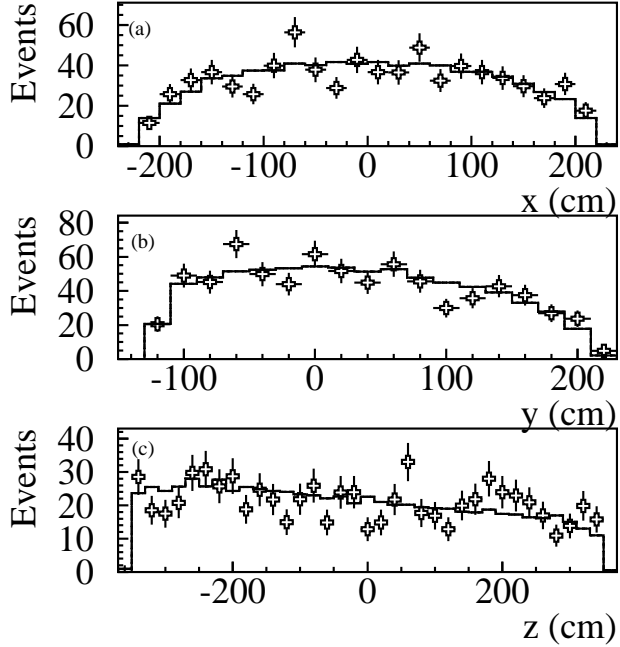,height=4in}}
\caption{The spatial distribution of the electron for beam-excess events 
compared with expectation (solid line) 
from $^{12}C(\nu_e,e^-)^{12}N_{g.s.}$.}
\label{fig:xyz}
\end{figure}

The reaction $\nu_e+~^{12}C\rightarrow e^-+~^{12}N_{g.s.}$ is identified 
by the detection of the $e^-$, followed within 60 ms by the positron 
from the $\beta$-decay of the $^{12}N_{g.s.}$.
Transitions to excited states of $^{12}N$ decay by prompt proton emission 
and thus do not feed down to the $^{12}N$ ground state or contribute
to the delayed coincidence rate.
The scattered electron has a maximum kinetic energy of 35.5 MeV 
due to the $Q$ value of 17.33 MeV. 
The $\beta$-decay has a mean lifetime of 15.9 ms and 
maximum positron kinetic energy of 16.33 MeV.
The cross section to the $^{12}N$ ground state has been calculated 
by several groups.
The form factors required to calculate the cross section are well known 
from a variety of previous measurements.
This cross section and the known $\nu_e$ flux are used 
to obtain the expected electron kinetic energy spectrum.   
Figure \ref{fig:eelec} shows the observed and expected electron energy 
distributions for events with an identified $\beta$-decay.
Figure \ref{fig:xyz} shows the observed and expected spatial 
distributions of the same events. 
Both figures show good agreement with expectations. 

\begin{figure}
\centerline{\psfig{figure=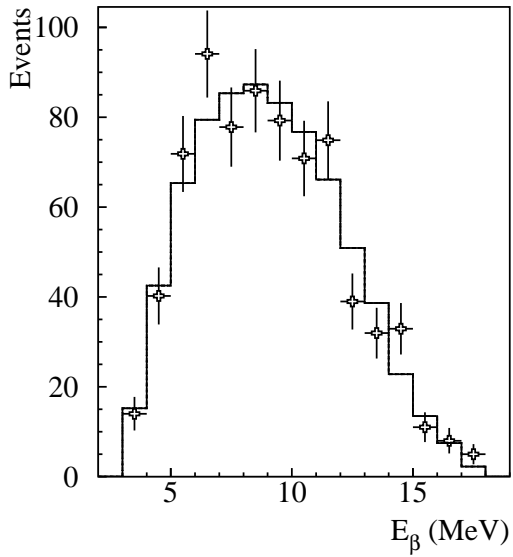,height=3.3in}}
\caption{Observed and expected (solid line) $e^+$ energy distribution 
for the $^{12}C(\nu_e,e^-)^{12}N_{g.s.}$ sample.}
\label{fig:epos}
\end{figure}

\begin{table}
\centering
\caption{Criteria to select $e^+$ from $N_{g.s.}$ beta decay and 
corresponding 
efficiencies for the reaction $^{12}C(\nu_e,e^-)^{12}N_{g.s.}$.}
\begin{tabular}{ccc}
\hline
Quantity            & Criteria            &    Efficiency   \\
\hline
$\beta$ decay time  & 52 $\mu$s$<t<60$ ms & 0.974$\pm$0.002 \\ 
Spatial correlation & $\Delta r<0.7$ m    & 0.992$\pm$0.008 \\
PMT threshold       & $>100$ for 1994,    & 0.856$\pm$0.011 \\
                    & $>75$ after 1994    &                 \\
Fiducial volume     & $D>0$ cm            & 0.986$\pm$0.010 \\
Trigger veto        & $>15.1~\mu$s        & 0.760$\pm$0.010 \\
In-time veto        & $<4$ PMTs           & 0.988$\pm$0.010 \\
DAQ dead time       &                     & 0.977$\pm$0.010 \\
\hline
Total               &                     & 0.598$\pm$0.015 \\ 
\hline
\end{tabular}
\label{ta:betaeff}
\end{table}
\begin{figure}
\centerline{\psfig{figure=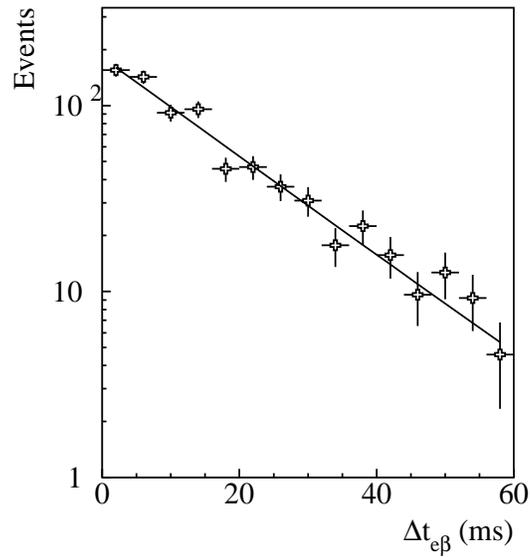,height=3.3in}}
\caption{Distribution of time between the $e^-$ and $e^+$ 
for beam excess events in the $^{12}C(\nu_e,e^-)^{12}N_{g.s.}$ sample.
The best fit curve (solid line) corresponds to 
a lifetime of $16.3\pm0.8$ ms.}
\label{fig:beta_dt}
\end{figure}
\begin{figure}
\centerline{\psfig{figure=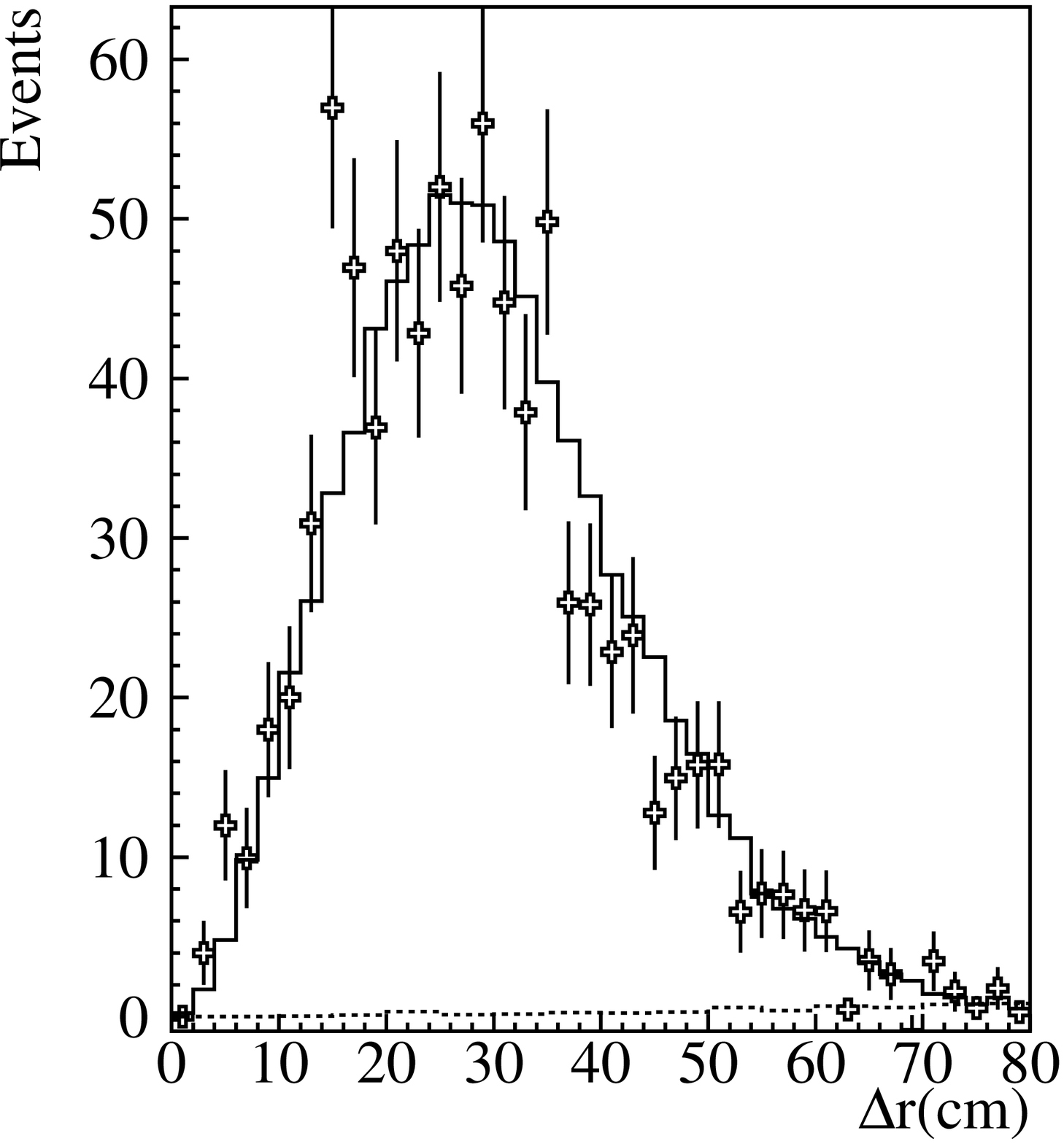,height=3.3in}}
\caption{Distribution of the distance between the reconstructed positions 
of the $e^-$ and $e^+$ for beam excess events 
in the $^{12}C(\nu_e,e^-)^{12}N_{g.s.}$ sample
compared with Monte Carlo expectation (solid line). 
The calculated accidental contribution is shown by the dashed line.}
\label{fig:beta_dr}
\end{figure}
Table \ref{ta:betaeff} gives the selection criteria and efficiencies 
for the $^{12}N~\beta$-decay positron.
Figure \ref{fig:beta_dt} shows the observed $\beta$-decay time 
distribution. 
The best fit curve yields a lifetime of $16.3\pm0.8$ ms in good agreement 
with the expected value of 15.9 ms.
Figure \ref{fig:beta_dr} shows the distance between the reconstructed 
electron and positron positions for the beam-excess sample. 
A cut was applied at 0.7 m, resulting in an acceptance 
of $(99.2\pm0.8)\%$. 
Following an electron produced by a neutrino interaction, 
an uncorrelated particle, such as the positron from $^{12}B~\beta$-decay, 
will occasionally satisfy all the positron criteria including 
the requirements 
of time (60 ms) and spatial (0.7 m) correlation with the electron. 
The probability of such an accidental coincidence was measured 
by using the sample of Michel electrons. 
The inefficiency caused by the 15.1 $\mu$s veto and the DAQ dead time 
are the same as for the electron.
Positrons with 4 or more in-time veto hits or any bottom veto coincidence 
are rejected. 
The energy distribution of the positron is calculated 
from the $^{12}N~\beta$-decay using
\begin{equation}
\frac{dN}{dE_e}=P_eE_e(E_{max}-E_e)^2\times\frac{2\pi\eta}{(e^{2\pi\eta}-1)}
\end{equation}
\noindent
where $\eta=Z\alpha/\beta_e$ and $E_e$ is the total positron energy 
(including rest energy). 
The slight modification of the spectrum due to the shape 
correction factor\cite{Kaina77} was found to have a negligible effect 
on the results.
The $^{12}N$ decays to the ground state ($E_{max}=16.83$ MeV) $94.6\%$ 
of the time. 
Beta decay transitions to the excited states of carbon are $1.9\%$ 
($E_{max}=12.39$ MeV, followed by a 4.44 MeV $\gamma$), 
$2.7\%~(E_{max}$ = 9.17 MeV) and $0.8\% (E_{max}$ = 6.5 MeV)\cite{FAS90}. 
The positron annihilates with an electron after stopping.
The Monte Carlo was used to generated expected distributions 
for the positron energy and for number of hit PMTs.  
There was a trigger requirement of 100 PMT hits for 1994 data and 
75 PMT hits after 1994.
The beta is required to be less than 18 MeV in this analysis.
Figure \ref{fig:epos} compares the observed and expected positron 
energy distributions. 
The good agreement shows that the energy calibration
is valid for these low energy electrons.

The electron and $\beta$ selection efficiencies,
the efficiency for satisfying the electron energy requirement 
and the total efficiency for the 
reaction $^{12}C(\nu_e,e^-)^{12}N_{g.s.}$ 
are shown in Table \ref{ta:gseff}.
\begin{table}
\centering
\caption{The electron and $\beta$ selection efficiencies, the efficiency 
for satisfying the electron energy requirement and the total efficiency 
for the process $^{12}C(\nu_e,e^-)^{12}N_{g.s.}$}
\begin{tabular}{cc}
\hline
Quantity             &     Efficiency      \\
\hline
Electron Selection   &   $0.492\pm0.035$   \\
Future $\beta$       &   $0.598\pm0.015$   \\
18 MeV $<E_e<40$MeV  &   $0.789\pm0.020$   \\
\hline
Total Efficiency     &   $0.232\pm0.019$   \\
\hline
\end{tabular}
\label{ta:gseff}
\end{table}

\begin{table}
\centering
\caption{Events, efficiency, neutrino flux and flux averaged cross 
section with statistical error only for $^{12}C(\nu_e,e^-)^{12}N_{g.s.}$}
\begin{tabular}{cc}
\hline
Beam-on events                               &      743        \\
Beam-off events                              &        6        \\
Accidental background                        &        4        \\
\hline
Events from $^{12}C(\nu_e,e^-)^{12}N_{g.s.}$ &      733        \\
Efficiency                                   & $0.232\pm0.019$ \\
$\nu_e$ flux                       & $10.58\times10^{13}$ cm$^{-2}$ \\
$\langle\sigma\rangle$             & $(8.9\pm0.3)\times10^{-42}$ cm$^2$ \\
\hline
\end{tabular}
\label{ta:gssample}
\end{table}
Table \ref{ta:gssample} provides a breakdown of the number of events 
satisfying the selection criteria as well as the total acceptance, 
the neutrino flux and the resulting flux averaged cross section. 
For the complete data sample the flux averaged cross section is 
$\langle\sigma\rangle=(8.9\pm0.3\pm0.9)\times10^{-42}$ cm$^2$ 
where the first error is statistical and the second is systematic. 
The two dominant sources of systematic error are the neutrino 
flux ($7\%$) discussed in Section \ref{sec:source} and the effective 
fiducial volume ($6\%$) discussed in Section \ref{sec:analysis}. 
The measured cross section decreases by $1.4\%$ 
when the fiducial volume is reduced by requiring that the electron be 
at least 50 cm (instead 35 cm) from the surface of the PMT faces.
As discussed in section \ref{sec:source} the beam stop was substantially 
modified after the 1995 run.
The cross section measured for data taken with the modified beam dump 
is $(5\pm6(stat.)~)\%$ higher than for the initial beam dump and 
thus fully consistent within the statistical uncertainty.
For comparison the previous measurements, the final LSND result 
and several theoretical predictions for the flux averaged cross section 
are presented in Table \ref{ta:xsec}. 
They are all in agreement with each other.
\begin{table}
\caption { Measurements and theoretical predictions of 
the flux averaged cross section for the process 
 $^{12}\rm{C}(\nu_e,e^-)^{12}\rm{N}_{g.s.}$.}
\label{ta:xsec}
\begin{tabular}{ll} 
\hline
Experiment                   &                                          \\
\hline
LSND                         & $(8.9\pm0.3\pm0.9)\times10^{-42}$ cm$^2$ \\
LSND(previous)\cite{LSND97a} & $(9.1\pm0.4\pm0.9)\times10^{-42}$ cm$^2$ \\
E225\cite{E225}              & $(10.5\pm1.0\pm1.0)\times10^{-42}$ cm$^2$\\
KARMEN\cite{KARMEN}          & $(9.1\pm0.5\pm0.8)\times10^{-42}$ cm$^2$ \\
\hline
\hline
Theory                                  &                              \\
\hline
Donnelly\cite{Donnelly}                 &  $9.4\times10^{-42}$ cm$^2$  \\ 
Fukugita {\it et al.}\cite{Fukugita88} &  $9.2\times10^{-42}$ cm$^2$  \\ 
Kolbe {\it et al.}\cite{Kolbe99}       &  $8.9\times10^{-42}$ cm$^2$  \\ 
Mintz {\it et al.}\cite{Mintz}         &  $8.0\times10^{-42}$ cm$^2$  \\ 
\hline
\end{tabular}
\end{table}
We note that all these measurements rely on the neutrino flux calibration 
that is discussed in Section \ref{sec:source} and 
thus they have correlated systematic errors.

\begin{figure}
\centerline{\psfig{figure=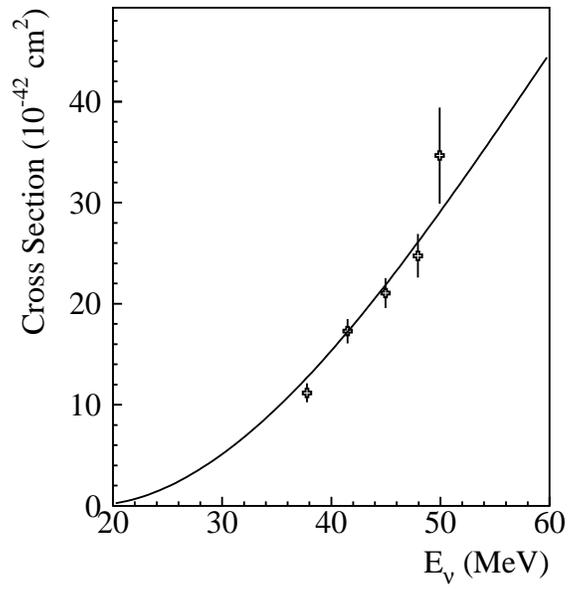,height=3.5in}}
\caption{The measured and expected (solid line) cross section 
for the process $^{12}C(\nu_e,e^-)^{12}N_{g.s.}$.}
\label{fig:xsec}
\end{figure}

\begin{figure}
\centerline{\psfig{figure=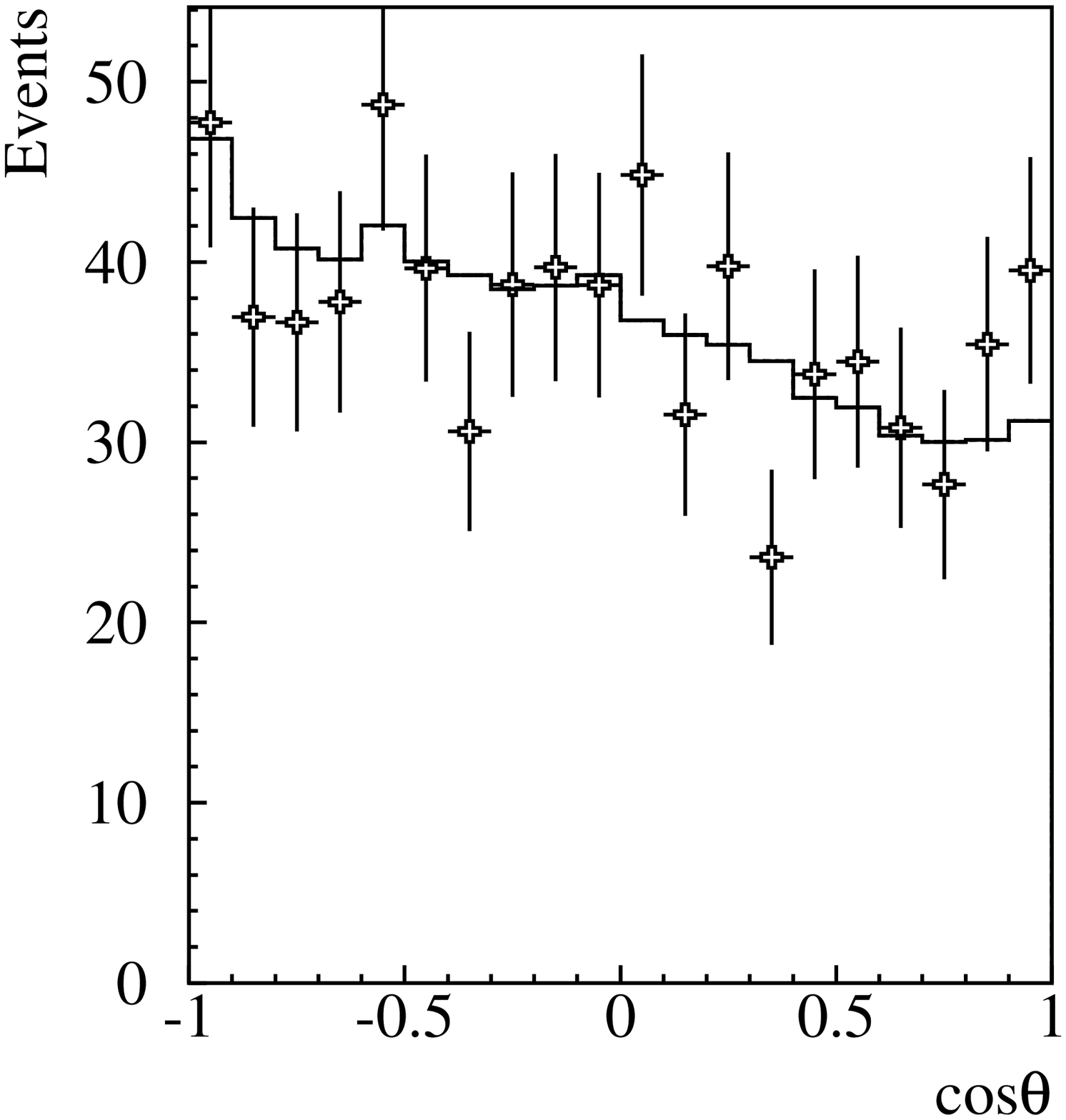,height=3.3in}}
\caption{Observed and expected (solid line) distribution in $\cos\theta$ 
for the $^{12}C(\nu_e,e^-)^{12}N_{g.s.}$ sample, 
where $\theta$ is the angle between the $e^-$ and the incident neutrino.}
\label{fig:angle}
\end{figure}

For this reaction to the $^{12}N$ ground state it is also 
straightforward to measure the energy dependence of the cross section.  
The recoil energy of the $^{12}N$ nucleus is negligible 
and thus $E_\nu=E_e+17.3$ MeV where $E_e$ is the electron kinetic energy.
Figure \ref{fig:xsec} shows that the energy dependence of the measured 
cross section agrees well with expectations\cite{Fukugita88}.
Figure \ref{fig:angle} shows the observed and 
expected\cite{Fukugita88,Kolbe96} 
angular distribution between the electron and the incident neutrino. 
The data agree well with expectations with the $\chi^2/DF=0.79$.
The mean observed value of $\cos\theta$ 
is $-0.046\pm0.021(stat.)\pm0.030(syst.)$ 
compared with an expected value of $-0.068$.
The systematic uncertainty in $\cos\theta$ is estimated to be 0.030 
based on Monte Carlo studies of the detector response.
The expected $\cos\theta$ distribution shown in Figure \ref{fig:angle} 
includes the effects of experimental resolution and as a result, is 
less backward peaked than the theoretical distribution used 
to generate it.

\section{Transitions to excited states of $^{12}N$}
\label{sec:exstate}

\begin{figure}
\centerline{\psfig{figure=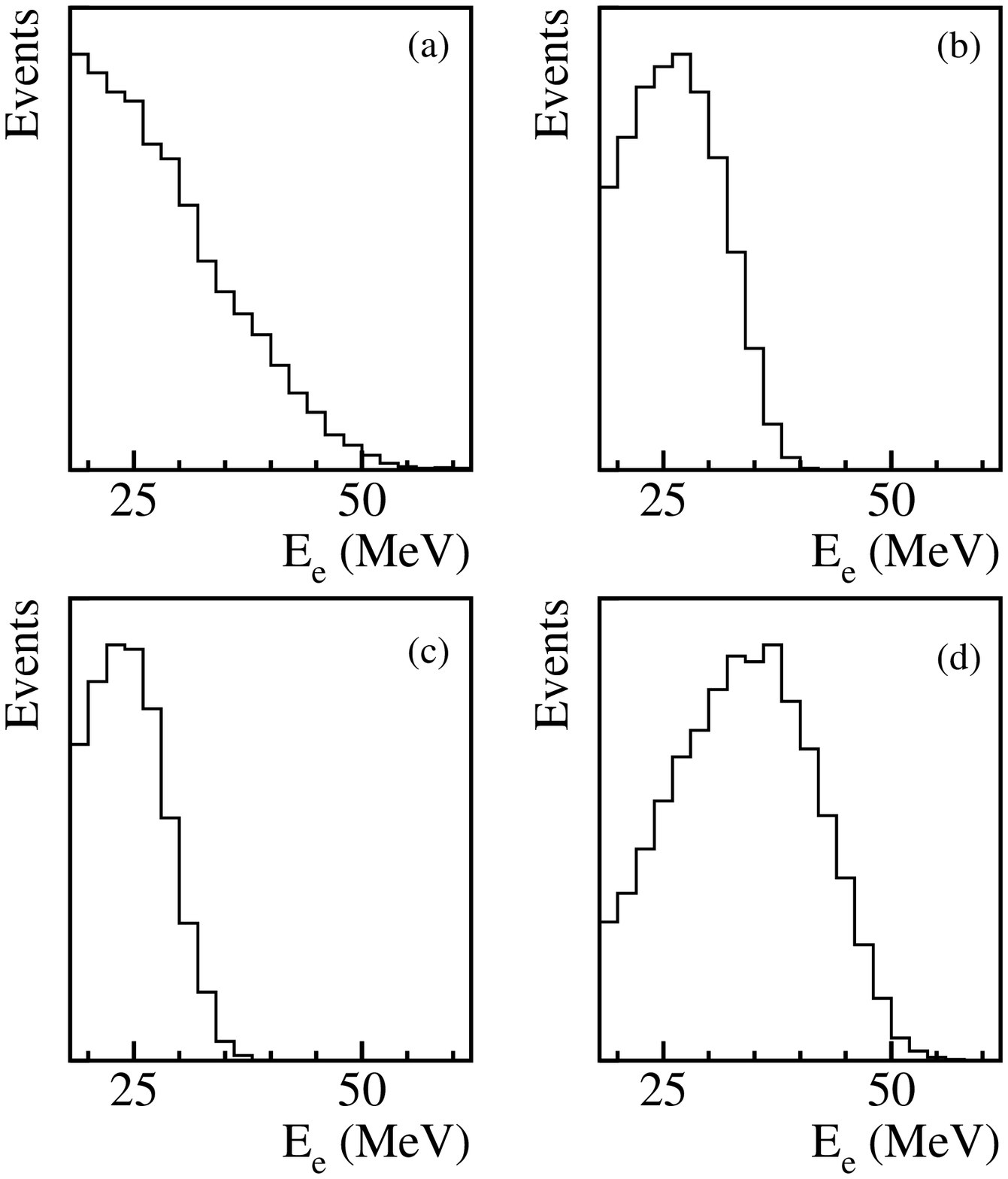,height=3.8in}}
\caption{The expected distributions of electron energy for 
(a) $\nu e^-$ elastic scattering, (b) $^{12}C(\nu_e,e^-)^{12}N_{g.s.}$, 
(c) $^{12}C(\nu_e,e^-)^{12}N^*$, and (d) $^{13}C(\nu_e,e^-)^{13}X$.}
\label{fig:pleemc}
\end{figure}

\begin{figure}
\centerline{\psfig{figure=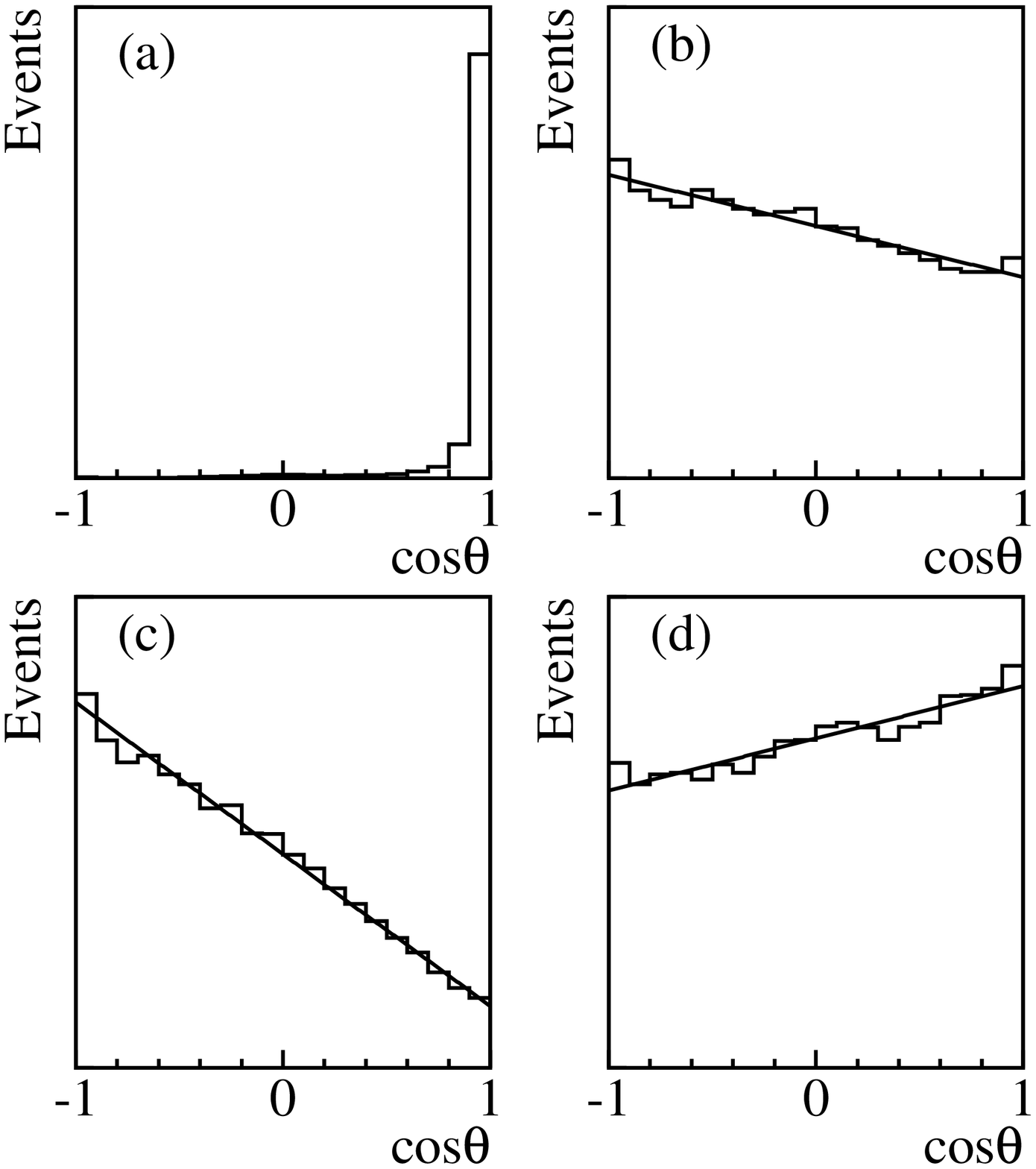,height=3.8in}}
\caption{The expected distributions of $\cos\theta$ for 
(a) $\nu e^-$ elastic scattering, (b) $^{12}C(\nu_e,e^-)^{12}N_{g.s.}$, 
(c) $^{12}C(\nu_e,e^-)^{12}N^*$, and (d) $^{13}C(\nu_e,e^-)^{13}X$.
Straight line fits are shown in (b), (c) and (d).}
\label{fig:plcos_mc}
\end{figure}

Electrons below 52 MeV are expected to arise from four major neutrino 
processes: $^{12}\rm{C}(\nu_e, e^-)^{12}\rm{N}_{g.s.}$, 
$^{12}\rm{C}(\nu_e, e^-)^{12}\rm{N}^*$, 
$^{13}\rm{C}(\nu_e, e^-)^{13}\rm{X}$ and 
neutrino electron elastic scattering.
The expected energy and angular distributions of these processes are 
shown in Figures \ref{fig:pleemc} and \ref{fig:plcos_mc}, respectively.  
The different event characteristics of these reactions are used to select 
a sample due primarily to the reaction $^{12}C(\nu_e,e^-)^{12}N^*$.
This sample is then used to determine the flux averaged cross section 
and the electron energy and angular distributions for this reaction.
The measured distributions are compared with expected distributions 
for $^{12}C(\nu_e,e^-)^{12}N^*$ obtained from LSNDMC 
using the theoretical cross section of Ref. \cite{Kolbe96}. 

The reaction $^{12}C(\nu_e,e^-)^{12}N_{g.s.}$ is  a source of background 
since the $e^+$ from the $\beta$-decay of $^{12}N_{g.s.}$ is not always 
identified.
Any event with an identified $e^+$ in delayed coincidence is of course 
excluded.  
The background of events with an unidentified $e^+$ is calculated using 
the positron acceptance given in Table \ref{ta:betaeff} and subtracted.

All three types of DAR neutrinos ($\nu_e,\nu_\mu$ and $\bar{\nu}_\mu$) 
elastically scatter off electrons in the detector 
but the rate is dominated by $\nu_ee^-$ scattering.
The contribution due to DIF $\nu_\mu$ and $\bar{\nu}_\mu$ scattering 
on electrons is almost negligible.
The scattered electron for this process is strongly forward peaked as 
shown in Figure \ref{fig:plcos_mc}, and thus such events can largely be 
eliminated with the requirement $\cos\theta<0.9$.
A measurement of electron-neutrino electron elastic scattering by LSND 
is reported in a separate paper\cite{LSND_nuee}.

A third background arises from the interaction of $\nu_e$ on $^{13}C$ 
nuclei ($1.1\%$ of the carbon). 
The expected number of events obtained from the calculated 
cross section\cite{Donnelly,Fukugita90} for this process is fairly small.
The $Q$ value is 2.1 MeV and thus about half of the background 
can be eliminated by requiring an electron energy below 34 MeV.  
We use the cross section calculated by Kubodera\cite{Fukugita90}, 
$0.525\times10^{-40}\rm{cm}^2$, and 
conservatively assign a $50\%$ uncertainty to this number.
The KARMEN experiment has measured a cross section of 
($0.5\pm0.37\pm0.1)\times10^{-40}$ cm$^2$ 
for this reaction\cite{KARMEN_xsec}.
%For this analysis the electron energy is required to be 
%between 18 and 34 MeV. 
%This region contains $68.7\%$ of the excited state events simulated 
%with LSNDMC using the cross section of Ref. \cite{Kolbe96}.

Most of the excited states of $^{12}N$ decay by prompt proton emission 
to $^{11}C$ states and the decays of some of these $^{11}C$ states 
produce gammas between 2 and 7 MeV.
Both these protons and gammas contribute to the visible energy detected 
by LSND and thus increase the apparent electron energy.
LSND is relatively insensitive to low energy protons, partly due to 
the absence of \v{C}erenkov light.
For example a 10 MeV proton produces only between 1 and 2 MeV electron 
equivalent energy.
The sensitivity to low energy gammas is also somewhat lower 
than for electrons of the same energy.
The response to protons, gammas and electrons is obtained using events 
simulated with LSNDMC.
For the analysis we require that the measured electron energy be 
between 18 and 34 MeV.	
Including the estimated contributions of protons and gammas 
based on Ref. \cite{Kolbe96} this region contains $72\pm4$\% 
of the excited state events.
The relatively large error assigned is due to uncertainties 
in the response of LSND, for example to low energy protons, as well as 
estimated uncertainties in the modeling of particle production.
The upper limit of 34 MeV not only eliminates much of 
the $^{13}C(\nu_e,e^-)^{13}X$ background, it further decreases 
the small background from the possible oscillation signal 
seen by LSND\cite{At96,At98}.

Table \ref{ta:exeff} shows the efficiency for each of the selection 
criteria and the total efficiency 
for the process $^{12}C(\nu_e,e^-)^{12}N^*$.
\begin{table}
\centering
\caption{The electron selection efficiency, the efficiency for satisfying 
the $\beta$ rejection criteria, the efficiencies for satisfying 
the angular and the energy requirements and the total efficiency 
for the process $^{12}C(\nu_e,e^-)^{12}N^*$}
\begin{tabular}{cc}
\hline
Quantity              &  Efficiency \\
\hline
Electron Selection    & 0.497$\pm$0.035 \\
No Future $\beta$     & 0.997$\pm$0.001 \\
$\cos\theta<0.9$      & 0.980$\pm$0.005 \\
18 MeV $<E_e<$ 34 MeV & 0.720$\pm$0.040 \\
\hline
Total Efficiency      & 0.349$\pm$0.031 \\
\hline
\end{tabular}
\label{ta:exeff}
\end{table}
Table \ref{ta:background} shows the calculated number of background 
events from various sources satisfying these selection criteria.
\begin{table}
\centering
\caption{Calculated number of background events 
satisfying the selection criteria for $^{12}C(\nu_e,e^-)^{12}N^*$.
Systematic uncertainties in the numbers of background events 
are also shown.}
\begin{tabular}{cc}
\hline
Source                              &  Events         \\
\hline
$^{12}C(\nu_e,e^-)^{12}N_{g.s.}$    & $434.1\pm28.1$ \\
$\nu e\rightarrow\nu e$             &  $35.2\pm4.1$  \\
$^{13}C(\nu_e,e^-)^{13}N$           &  $46.6\pm23.3$ \\
$^{12}C(\nu,\nu)^{12}C^*$[15.1 MeV] &  $17.5\pm2.0$  \\
$^{12}C(\nu_\mu,\mu^-)^{12}N^*$     &   $9.3\pm2.0$  \\
Events with a neutron               &  $21.5\pm15.1$ \\
\hline
Total Background                    &  $564\pm40$ \\
\hline
\end{tabular}
\label{ta:background}
\end{table}
In addition to the background sources already discussed, 
there are several smaller sources of background.

The neutral current 
excitation $^{12}C(\nu,\nu)^{12}C^*~(1^+$,1; 15.1 MeV) 
leads to prompt decay to photons with a $90\%$ branching ratio.
Most of the photons are eliminated by the 18 MeV energy requirement 
but approximately $0.8\%$ will have reconstructed energies above 18 MeV 
due to the finite energy resolution.
The measured cross section for this process\cite{Bodmann91,Armbruster98} 
is in good agreement with theoretical 
calculations\cite{Fukugita88,Donnelly,Kolbe94}.

As discussed earlier, LSND has also measured 
the process $^{12}C(\nu_\mu,\mu^-)^{12}N^*$ using a beam 
of $\nu_\mu$ from $\pi^+$ DIF\cite{LSND97b}.
Normally we detect both the $\mu^-$ and the Michel $e^-$ 
from the $\mu^-$ decay.
Occasionally, however, we will miss a low energy $\mu^-$ 
because it does not satisfy the activity threshold of $>17$ PMT hits.
In that case the Michel electron will be a background event 
if it satisfies the selection criteria given in Table \ref{ta:exeff}.
We have calculated this background 
using the event simulation program LSNDMC.
We have also estimated this background 
using the observed distribution of PMT 
hits for $\mu^-$ satisfying the activity threshold.

Finally, we determine the background from processes 
with associated neutrons.
Events from the reactions $\bar{\nu}_e+p\rightarrow e^++n$ 
and $\bar{\nu}_\mu+p\rightarrow\mu^++n$ are identified in LSND 
by detecting the 2.2 MeV $\gamma$'s from the capture 
reaction $n+p\rightarrow d+\gamma$\cite{LSND97a,LSND97b,At96}.
The mean capture time in the LSND detector is 186 $\mu$s, 
essentially independent of the initial neutron energy.
Three variables are used to identify a capture $\gamma$ correlated 
with a neutron in the primary event:
the number of PMT hits for the $\gamma$,
the distance of the $\gamma$ from the primary event and 
the time of $\gamma$ from the primary event.
A likelihood technique, discussed in Ref. \cite{At96}, has been developed 
to separate the correlated component due to neutrons 
from the uncorrelated or accidental component.
An approximate likelihood ratio $R={\cal L}_{cor}/{\cal L}_{uncor}$ is 
calculated for each event from the three measured variables.
We use here an analysis with improved photon spatial reconstruction as
is presented in Ref. \cite{LSND_osc}, which reports final LSND results 
on neutrino oscillations.

\begin{figure}
\centerline{\psfig{figure=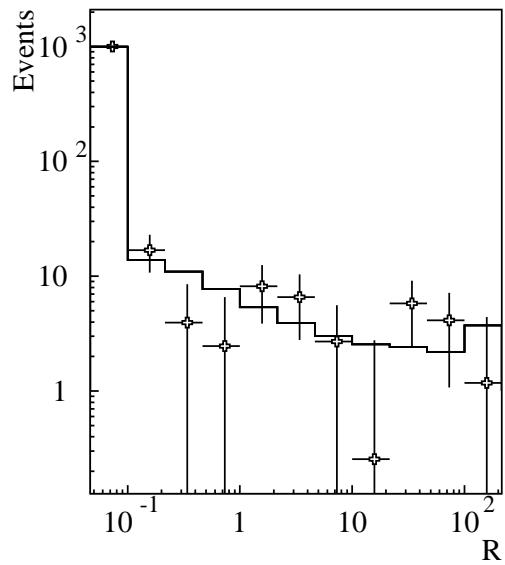,height=3.3in}}
\caption{The observed distribution of the $\gamma$ likelihood ratio R 
for the process $^{12}C(\nu_e,e^-)^{12}N^*$.
Shown for comparison is the best fit (solid line) combination 
of the correlated distribution and the uncorrelated distribution to data. 
The best fit has a $2.0\pm1.4\%$ correlated component.}
\label{fig:r_ex}
\end{figure}

Figure \ref{fig:r_ex} shows the observed $R$ distribution for events 
satisfying the selection criteria 
for the process $^{12}C(\nu_e,e^-)^{12}N^*$.
Shown for comparison is
the best fit (solid line) combination of the correlated distribution 
and the uncorrelated distribution to the data.
The best fit has a $2.0\pm1.4\%$ correlated component.
The number of background events with neutrons shown 
in Table \ref{ta:background} is obtained from the best fit to the data.

Table \ref{ta:ex_xsec} shows the number of beam-on and beam-excess events 
satisfying the selection criteria, the number of background events and 
the resulting number of events and cross section 
for the process $^{12}C(\nu_e,e^-)^{12}N^*$.
\begin{table}
\centering
\caption{Events, efficiency, neutrino flux and flux averaged cross 
section with statistical errors only for $^{12}C(\nu_e,e^-)^{12}N^*$}
\begin{tabular}{cc}
\hline
Beam-on events              &              2126                  \\
Beam-excess events          &              1088                  \\
Background                  &               564                  \\
\hline
$^{12}C(\nu_e,e^-)^{12}N^*$ &               524                  \\
Efficiency                  & $0.349\pm0.031$                    \\
Flux                        & $10.58\times10^{13}$ cm$^{-2}$     \\
$\langle\sigma\rangle$      & $(4.3\pm0.4)\times10^{-42}$ cm$^2$ \\
\hline
\end{tabular}
\label{ta:ex_xsec}
\end{table}
It is worth noting that the beam-off subtraction is much larger 
in this case than it was for the exclusive reaction 
where we require a coincidence with a $\beta$.
The flux average cross section 
is $\langle\sigma\rangle=(4.3\pm0.4\pm0.6)\times10^{-42}$ cm$^2$.
There are several contributions to the systematic error.
The $7\%$ flux uncertainty and the $6\%$ uncertainty in the effective 
fiducial volume have been described previously.
There is a $4\%$ uncertainty due to the $50\%$ error 
in the $^{13}C$ cross section. 
The uncertainty in the $e^+$ acceptance 
for the $^{12}N_{g.s.}$ background 
subtraction leads to a $5\%$ uncertainty in the $^{12}N^*$ cross section.
The uncertainty in the duty ratio results in a $3\%$ error 
in the cross section.
We estimate a $5.5\%$ uncertainty in the fraction of events 
with electrons in the region 18 MeV$<E_e<34$ MeV.

The cross section reported here is lower than that previously measured 
by LSND\cite{LSND97a} primarily because of an increase in the calculated 
number of background events. (see Table \ref{ta:background}).
An increase in the calculated deadtime due to the veto lowered 
the $\beta$ selection efficiency and increased the calculated number 
of events from the process $^{12}C(\nu_e,e^-)^{12}N_{g.s.}$ 
without an identified $\beta$.
There was also an increase in the calculated number of events 
with $\cos\theta<0.9$ from the process $\nu e\rightarrow\nu e$.
The small DIF background from $^{12}C(\nu_\mu,\mu^-)^{12}N^*$ was 
not previously included.
Finally, as was stated in Ref \cite{LSND97a}, 
we previously did not subtract the background from events with neutrons.

The flux averaged cross section measured by LSND is compared 
in Table \ref{ta:exxsec} with other measurements and with several 
theoretical calculations.
\begin{table}
\caption { Measurements and theoretical predictions of the flux averaged 
cross section for the process  $^{12}C(\nu_e,e^-)^{12}N^*$.}
\label{ta:exxsec}
\begin{tabular}{ll} 
\hline
Experiment                   &                                          \\
\hline
LSND                         & $(4.3\pm0.4\pm0.6)\times10^{-42}$ cm$^2$ \\
LSND(previous)\cite{LSND97a} & $(5.7\pm0.6\pm0.6)\times10^{-42}$ cm$^2$ \\
E225\cite{E225,E225_note}    & $(3.6\pm2.0)\times10^{-42}$ cm$^2$       \\
KARMEN\cite{KARMEN_xsec}     & $(5.1\pm0.6\pm0.5)\times10^{-42}$ cm$^2$ \\
\hline
\hline 
Theory                            &                                \\
\hline 
Kolbe {\it et al.}\cite{Kolbe94} &   $6.3\times10^{-42}$ cm$^2$   \\ 
Kolbe {\it et al.}\cite{Kolbe99} &   $5.5\times10^{-42}$ cm$^2$   \\ 
Hayes {\it et al.}\cite{Hayes00} &   $4.1\times10^{-42}$ cm$^2$   \\
\hline
\end{tabular}
\end{table}
The most recent measurements of both LSND and KARMEN\cite{KARMEN_xsec} 
are somewhat lower than their previous measurements\cite{KARMEN,LSND97a}.
The recent CRPA result of Kolbe {\it et al.}\cite{Kolbe99}, similarly, 
is lower than the previous result\cite{Kolbe94} but 
it remains above the cross section measurements of both LSND and KARMEN.
In contrast the recent shell model calculation of Hayes and 
Towner\cite{Hayes00} is in good agreement with but lower 
than the cross section measured by LSND.

\begin{figure}
\centerline{\psfig{figure=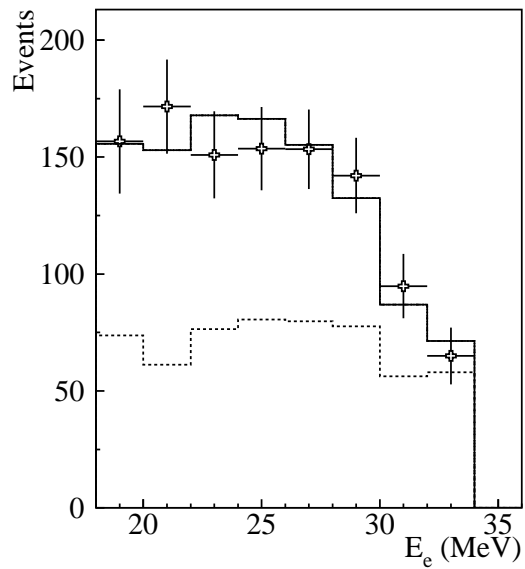,height=3.3in}}
\caption{The observed distribution of electron energy for events 
satisfying the selection criteria given in Table \ref{ta:exeff}.
The dotted (solid) histogram shows the expected distribution 
from all sources (background sources only).}
\label{fig:ee1}
\end{figure}

\begin{figure}
\centerline{\psfig{figure=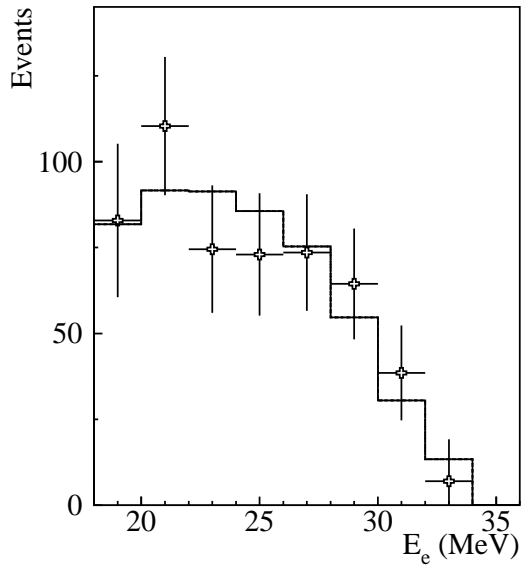,height=3.3in}}
\caption{The observed and expected (solid line) electron energy 
distribution for the process $^{12}C(\nu_e,e^-)^{12}N^*$}
\label{fig:ee2}
\end{figure}

Figure \ref{fig:ee1} shows that the measured electron energy distribution 
for the sample of events satisfying the electron criteria given 
in Table \ref{ta:exeff} agrees well with that expected 
from $^{12}C(\nu_e,e^-)^{12}N^*$ plus the background sources listed 
in Table \ref{ta:background}.
Also shown is the energy distribution expected from just the background 
processes.
For the largest background source, $^{12}C(\nu_e,e^-)^{12}N_{g.s.}$, 
we used the shape of the energy distribution measured 
for the sample of events with identified $\beta$'s. 
For all other processes the expected energy distributions are obtained 
using simulated samples of events. 
Figure \ref{fig:ee2} compares the measured energy distribution 
after subtraction of the calculated backgrounds with the distribution 
expected for the process $^{12}C(\nu_e,e^-)^{12}N^*$.
The agreement is excellent.

\begin{figure}
\centerline{\psfig{figure=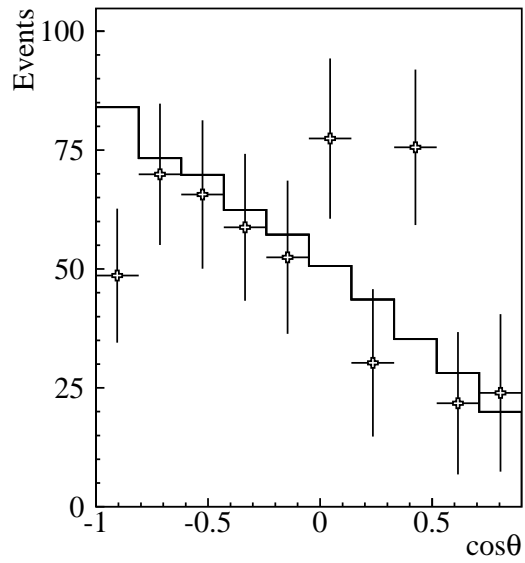,height=3.3in}}
\caption{The observed and expected (solid line) $\cos\theta$ distribution 
for the process $^{12}C(\nu_e,e^-)^{12}N^*$}
\label{fig:cos2}
\end{figure}

Figure \ref{fig:cos2} compares the measured distribution of $\cos\theta$ 
after subtraction of the calculated backgrounds with the distribution 
expected for the process $^{12}C(\nu_e,e^-)^{12}N^*$.
The $\cos\theta$ distribution is enhanced in the backward direction 
as expected.
The data are in fair agreement with expectations\cite{Kolbe96} 
with the $\chi^2/DF=1.79$.
The mean observed value of $\cos\theta$ 
is $-0.15\pm0.05(stat.)\pm0.030(syst.)$ compared with the expected value 
of $-0.25$ for Ref. \cite{Kolbe96} and $-0.30$ for Ref. \cite{Hayes00}.
The backward peaking of the angular distribution is largely a result 
of the negative parity of the $N^*$ states expected to contribute, 
$2^-$ levels at 1.20 and 4.14 MeV and $1^-$ levels at 6.40 and 7.68 MeV.
The $l=1$ angular momentum transfer to the $A=12$ system favors 
momentum transfer of approximately 100 MeV/c, and 
hence the backward peaking.

The total charged current cross section for $\nu_e$ interactions 
on $^{12}C$ is obtained by adding the cross sections measured 
for $^{12}C(\nu_e,e^-)^{12}N_{g.s.}$ and $^{12}C(\nu_e,e^-)^{12}N^*$. 
The resulting flux averaged cross section for  $^{12}C(\nu_e,e^-)^{12}N$ 
is $\langle\sigma\rangle=(13.2\pm0.5\pm1.3)\times10^{-42}$ cm$^2$.

\section{Conclusions}

The process $^{12}C(\nu_e,e^-)^{12}N_{g.s.}$ has been measured with 
a clean sample of 733 events.
For this process the cross section calculations 
using empirical form factors are expected to be very reliable.
The flux averaged cross section is measured to 
be $(8.9\pm0.3\pm0.9)\times10^{-42}$ cm$^2$ in reasonable agreement 
with other experiments and theoretical expectations.
The energy and angular distributions also agree well 
with theoretical expectations.

The measurement of the process $^{12}C(\nu_e,e^-)^{12}N^*$ is 
more difficult, 
primarily due to the significant background subtraction required.
The measured cross section of $(4.3\pm0.4\pm0.6)\times10^{-42}$ cm$^2$ is 
in agreement with other measurements.
It is in better agreement with the shell model calculation of Hayes and 
Towner \cite{Hayes00} than with the CRPA calculation 
of Kolbe {\it et. al.}\cite{Kolbe99}, but is compatible with both models.

~~~~~

{\it Acknowledgments} 
This work was conducted under the auspices of the US Department 
of Energy, supported in part by funds provided by the University 
of California for the conduct of discretionary research 
by Los Alamos National Laboratory.
This work was also supported by the National Science Foundation.
We are particularly grateful for the extra effort that was made 
by these organizations to provide funds for running the accelerator 
at the end of the data taking period in 1995.
It is pleasing that a number of undergraduate students from participating 
institutions were able to contribute significantly to the experiment.

\clearpage


\begin{thebibliography}{99}
\bibitem{E225_note} The quoted value of $3.6\times10^{-42}$ cm$^2$, resulting from the subtraction of the ground state cross section from the inclusive cross section, is the preferred number of two values given in ref \cite{E225}.
\end{thebibliography}
\end{document}